\documentclass[twocolumn,trackchanges]{aastex631}
\usepackage{booktabs} 
\usepackage{amsmath, amssymb, amsfonts}
\usepackage{soul} 
\usepackage{multirow}
\usepackage{threeparttable}
\usepackage{subfigure}
\usepackage{url}
\usepackage{bm}
\graphicspath{{./figure}}

\newcommand\appendixfootnote[1]{\footnote{\hsize=\columnwidth\advance\hsize by-18pt\relax#1}}

\begin{document}
    \title{A constraint on superheavy elements of the GRB-kilonova AT 2023vfi}
    \correspondingauthor{Zhengyan Liu, Wen Zhao}
    \email{ustclzy@mail.ustc.edu.cn, wzhao7@ustc.edu.cn}

    \author[0000-0002-2242-1514]{Zhengyan Liu}
    \affiliation{CAS Key Laboratory for Researches in Galaxies and Cosmology, Department of Astronomy, University of Science and Technology of China, Chinese Academy of Sciences, Hefei, Anhui 230026, China}
    \affiliation{School of Astronomy and Space Sciences, University of Science and Technology of China, Hefei 230026, China}

    \author[0000-0002-9092-0593]{Ji-an Jiang}
    \affiliation{CAS Key Laboratory for Researches in Galaxies and Cosmology, Department of Astronomy, University of Science and Technology of China, Chinese Academy of Sciences, Hefei, Anhui 230026, China}
    \affiliation{School of Astronomy and Space Sciences, University of Science and Technology of China, Hefei 230026, China}
    \affiliation{National Astronomical Observatory of Japan, 2-21-1 Osawa, Mitaka, Tokyo 181-8588, Japan}
    
    \author[0000-0002-1330-2329]{Wen Zhao}
    \affiliation{CAS Key Laboratory for Researches in Galaxies and Cosmology, Department of Astronomy, University of Science and Technology of China, Chinese Academy of Sciences, Hefei, Anhui 230026, China}
    \affiliation{School of Astronomy and Space Sciences, University of Science and Technology of China, Hefei 230026, China}
    
    \begin{abstract}
        The discovery of the kilonova (KN) AT 2017gfo, accompanying the gravitational wave event GW170817, provides crucial insight into the synthesis of heavy elements during binary neutron star (BNS) mergers. Following this landmark event, another KN was detected in association with the second-brightest gamma-ray burst (GRB) observed to date, GRB 230307A, and subsequently confirmed by observations of the James Webb Space Telescope (JWST). In this work, we conduct an end-to-end simulation to analyze the temporal evolution of the KN AT 2023vfi associated with GRB 230307A, and constrain the abundances of superheavy elements produced. We find that the temporal evolution of AT 2023vfi is similar to AT 2017gfo in the first week post-burst. Additionally, the \textit{r}-process nuclide abundances of lanthanide-rich ejecta, derived from numerical relativity simulations of BNS mergers, can also successfully interpret the temporal evolution of the KN with the lanthanide-rich ejecta mass of $0.02 M_\odot$, which is consistent with the mass range of dynamical ejecta from numerical simulations in literature. Both findings strongly suggest the hypothesis that GRB 230307A originated from a BNS merger, similar to AT 2017gfo. Based on the first time observation of the KN for JWST, we are able to constrain the superheavy elements of another KN following AT 2017gfo. The pre-radioactive-decay abundances of the superheavy nuclides: $^{222}$Rn, $^{223}$Ra, $^{224}$Ra and $^{225}$Ac, are estimated to be at least on the order of $1 \times 10^{-5}$. These abundance estimates provide valuable insight into the synthesis of superheavy elements in BNS mergers, contributing to our understanding of astrophysical \textit{r}-process nucleosynthesis.
    \end{abstract}
    
    \section{introduction} \label{sec_intro}
        Mergers of binary neutron stars (BNSs) and neutron star-black hole (NS-BH) binaries have been proposed as promising sites for rapid neutron capture (\textit{r}-process) nucleosynthesis \citep{lattimer1974,lattimer1977,eichler1989}. The radioactive decay of \textit{r}-process nuclei heats the expanding material ejected during these mergers, producing an electromagnetic transient in ultraviolet, optical, and near-infrared wavelengths, known as ``kilonova" (KN) \citep{Li_1998,Metzger_2010,metzger_kilonovae_2020}. The mergers of BNS and NS-BH also produce a relativistic jet, which is detected as a short-duration gamma-ray burst (sGRB) if the jet is oriented toward the observer \citep{pacynski1986,narayan1992,popham1999}, accompanied by a KN and a gravitational-wave (GW) burst \citep{Metzger_Berger_2012,Tanvir_2013,Troja_2019a,Jin_2019,Rastinejad_2021}. \par
        On 2017 August 17 12:41:04.47 UTC, the Advanced Laser Interferometer Gravitational-Wave Observatory (aLIGO) and Virgo detected the first BNS merger GW source, GW170817 \citep{abbott_gw170817_2017}. Shortly after the GW signal, the sGRB 170817A \citep{goldstein_ordinary_2017,savchenko_integral_2017} and the kilonova AT 2017gfo \citep{coulter_swope_2017,shappee_early_2017,cowperthwaite_electromagnetic_2017,evans_swift_2017} were observed $\sim1.7$ s and $\sim11$ hr after GW alert, respectively. Both were confirmed to be associated with this BNS merger event. According to the analysis of the kilonova AT 2017gfo, the ejecta consists of at least two distinct components, one composed mainly of light \textit{r}-process elements (atomic mass number $A\leq140$, lanthanide-poor), responsible for the early blue emission, and another of heavy elements ($A>140$, lanthanide-rich), which yields the later red emission of the kilonova \citep{kasen_origin_2017,Villar_2017,Nicholl_2017}. In general, the lanthanide-rich ejecta originates from dynamical ejecta, and the lanthanide-poor ejecta is composed of shock heated dynamical ejecta and disk outflow material \citep{metzger_kilonovae_2020}. The discovery of AT 2017gfo, associated with the GW event GW170817, provides smoking-gun evidence for the synthesis of heavy elements via the BNS merger. \par
        The lightcurve of a KN could be influenced by various energy sources, including energy released by radioactive decay of \textit{r}-process elements, shock cooling \citep{kasliwal_illuminating_2017,Piro_2018}, energy injection from the merger remnant \citep{Yu_2018,Kawaguchi_2020,Ai_2022,Ai_2024}, and interaction between ejecta and circumstellar medium \citep[CSM;][]{2022ApJ...925...43Q,2022ApJ...940L..44R,Wang_2024a}. Among these, the primary source is the energy released by radioactive decay. Several studies have investigated the connection between KN lightcurve and heavy elements \citep{Metzger_2010,Lippuner_2015,Wanajo_2018,Wu_2019,hotokezaka_radioactive_2020,Chen_2023a}. \cite{Metzger_2010} related the radioactive heating rate to KN lightcurves using a nuclear reaction network. A comprehensive parameter research by \cite{Lippuner_2015} explored the variability of heating rate with the richness of lanthanide element. \cite{hotokezaka_radioactive_2020} refined the modeling of the thermalization process to accurately derive the heating rate of \textit{r}-process elements. \cite{Wu_2019} identified the significant impact of energy released by the radioactive decay of certain superheavy nuclides (atomic mass number $A>210$) on KN lightcurves at later epochs, suggesting that such features could constrain the abundances of these nuclides. Moreover, according to this feature, \cite{Chen_2023a} used the lightcurve of AT 2017gfo to estimate the abundance of $^{225}$Ac in the kilonova ejecta. \par
        To better measure the abundances of these superheavy elements, obtaining the complete evolution of KN is essential. To date, a total of 17 KNe or KN candidates have been reported, including 15 KNe collected in \cite{Li_KN_2023} and two others reported in \cite{rastinejad2022kilonova} and \cite{Levan_2024}, respectively. The superheavy elements are difficult to be constrained for most KNe, due to the lack of late-time observation. With the high sensitivity in near-infrared wavelength, the James Webb Space Telescope (JWST) is an ideal and powerful instrument for following and monitoring KN evolution \citep{Bartos_2016,Wu_2019,Chen_2023a}. Recently, the second brightest GRB observed to date, GRB230307A was detected by several high-energy instruments \citep{2023GCN.33407....1D,2023GCN.33406....1X,Sun_2023,2023GCN.33413....1K,2023GCN.33427....1S}. Subsequent optical follow-ups revealed evidence of a KN component associated with this GRB, named as AT 2023vfi\footnote{\href{https://www.wis-tns.org/object/2023vfi}{https://www.wis-tns.org/object/2023vfi}}. Based on the KN spectral models \citep{Domoto_2025,2025MNRAS.536.2973P}, some spectral features of \textit{r}-process elements (e.g., strontium, tellurium, thorium) are expected to be observed. To confirm and further measure the late-time evolution of the KN, JWST was employed to follow up the KN for the first time \citep{Levan_2024}. According to subsequent spectral analyses \citep{Gillanders_2023,Levan_2024,gillanders_2024}, certain spectral features potentially attributable to \textit{r}-process elements (e.g., tellurium) were reported, similar to those observed in AT 2017gfo \citep{Hotokezaka_2023}, indicating the origin of BNS or NS-BH merger for GRB 230307A. \par
        Unlike most previously discovered KNe associated with sGRBs \citep{Li_KN_2023}, AT 2023vfi is associated with the long-duration GRB (lGRB) 230307A \citep{Levan_2024,Yang_2024}. The duration of a GRB is defined by $T_{90}$, the time interval during which the fluence of the GRB increases from 5\% to 95\%. Multi-wavelength observations suggest that sGRBs ($T_{90}<2$ s) and lGRBs ($T_{90}>2$ s) originate from compact star mergers and massive star core collapses \citep{Woosley_2006,Tanvir_2013,abbott_gw170817_2017}, respectively. However, the recent discovery of KNe associated with lGRBs (e.g., GRB 211211A, GRB 230307A) challenges this well-known duration-progenitor connection \citep{rastinejad2022kilonova,Levan_2024}. \cite{Zhang_2024a} suggests that the features of GRB 230307A shows a engine-defined long duration, which can be accommodated with a BNS merger. \par
        In this work, assuming a BNS merger origin, we focus on constraining the abundances of superheavy elements reported in \cite{Wu_2019} for the KN associated with GRB 230307A. We conduct detailed \textit{r}-process nucleosynthesis simulations and derive the corresponding KN lightcurves to compare with the observations. Additionally, we discuss other mechanisms that may influence the late-time KN lightcurve. The paper is organized as follows: In Section \ref{method}, we introduce the method to extract the KN lightcurve and describe \textit{r}-process nucleosynthesis simulations and KN lightcurve calculation. We present the simulated lightcurves and the estimated abundances of the superheavy elements and discuss the influences of other energy sources on our results in Section \ref{results}. Finally, we provide conclusions in Section \ref{summary}.\par
    \section{method} \label{method}
        \subsection{Extract the KN lightcurve} \label{method.1}
            Except for the KN AT 2017gfo, all KNe discovered so far are mixed with GRB afterglows, including AT 2023vfi. In this section, we introduce data usage and the reduction method of multi-wavelength data to extract the KN temporal evolution. \par
            The origins and characteristics of afterglows and KNe are distinct. Afterglows arise from the interaction between the relativistic jet and the surrounding interstellar medium, resulting in non-thermal synchrotron radiation across a broad range of wavelengths from X-ray to optical and radio. KNe are approximately thermal emissions in ultraviolet, optical, and near-infrared wavelengths, heated by the radioactive decay of \textit{r}-process elements formed during the BNS and NS-BH mergers. According to the difference in their spectral energy distribution (SED) shapes, the combination of multi-wavelength data (e.g., X-ray, optical) is commonly used to identify and separate KNe from afterglows \citep[e.g.,][]{Tanvir_2013,Jin_2019,Troja_2019a,Rastinejad_2021}. In this work, we adopt the same method used in \cite{Yang_2024} to extract the KN. The X-ray data used here is obtained from the Neil Gehrels Swift Observatory \citep[Swift;][]{Burrows_2005} and the High Throughput X-ray Spectroscopy Mission and the X-ray Multi-Mirror Mission \citep[XMM-Newton;][]{Jansen_2001}, as published in \cite{Yang_2024}. For optical and near-infrared data, we use the photometry results from various ground-based telescopes and JWST, as presented in \cite{Levan_2024} and \cite{Yang_2024}. \par
            \begin{table}[h]
                \centering
                \footnotesize
                \setlength{\tabcolsep}{4pt}
                \renewcommand{\arraystretch}{1.3}
                \caption{The ultraviolet, optical and near-infrared data used in SED fitting at different epochs. $\Delta t$ represents the time since the occurrence of GRB 230307A, in units of days. All errors represent the $1\sigma$ uncertainties. References: (1) \cite{Levan_2024}; (2) \cite{Yang_2024}; (3) \cite{2023GCN.33459....1B}}. 
                \begin{tabular}{*{6}{c}}
                \toprule 
                {$\Delta t$} & Magnitude & Filter & Telescope & Epoch & Ref\\
                \midrule
                1.12 & $20.90\pm0.05$ & R & KMTNet/SAAO & 1.2 & (2)\\
                1.20 & $20.21\pm0.15$ & H & PRIME & 1.2 & (2)\\
                1.20 & $20.74\pm0.11$ & J & PRIME & 1.2 & (2)\\
                1.20 & $20.60\pm0.14$ & Y & PRIME & 1.2 & (2)\\
                1.20 & $22.1\pm0.2$ & white & UVOT & 1.2 & (2)\\
                1.43 & $20.72\pm0.15$ & r & ULTRACAM & 1.2 & (1)\\
                1.60 & $23.0\pm0.3$ & white & UVOT & 1.8 & (2)\\
                1.82 & $21.20\pm0.10$ & I & KMTNet/SSO & 1.8 & (2)\\
                1.82 & $21.42\pm0.05$ & R & KMTNet/SSO & 1.8 & (2)\\
                2.36 & $21.50\pm0.12$ & z' & SOAR & 2.4 & (3)\\
                2.37 & $21.84\pm0.19$ & r & VST & 2.4 & (1)\\
                2.38 & $22.46\pm0.06$ & R & KMTNet/CTIO & 2.4 & (2)\\
                2.39 & $22.00\pm0.20$ & I & KMTNet/CTIO & 2.4 & (2)\\
                2.41 & $22.35\pm0.26$ & g & ULTRACAM & 2.4 & (1)\\
                2.41 & $21.68\pm0.09$ & i & ULTRACAM & 2.4 & (1)\\
                6.42 & $23.24\pm0.11$ & z & FORS2 & 7.4 & (1)\\
                7.40 & $24.90\pm0.10$ & r & Gemini & 7.4 & (2)\\
                8.34 & $24.30\pm0.20$ & z & Gemini & 7.4 & (2)\\
                28.83 & $28.50\pm0.07$ & F115W & JWST & 28.9 & (1)\\
                28.83 & $26.24\pm0.01$ & F277W & JWST & 28.9 & (1)\\
                28.86 & $28.11\pm0.12$ & F150W & JWST & 28.9 & (1)\\
                28.86 & $25.42\pm0.01$ & F356W & JWST & 28.9 & (1)\\
                28.89 & $28.97\pm0.20$ & F070W & JWST & 28.9 & (1)\\
                28.89 & $24.62\pm0.01$ & F444W & JWST & 28.9 & (1)\\
                61.48 & $29.78\pm0.31$ & F115W & JWST & 61.4 & (1)\\
                61.48 & $26.97\pm0.04$ & F444W & JWST & 61.4 & (1)\\  
                61.51 & $28.31\pm0.12$ & F277W & JWST & 61.4 & (1)\\ 
                61.51 & $29.24\pm0.17$ & F150W & JWST & 61.4 & (1)\\
                \bottomrule
                \end{tabular}
                \label{tab:UV-OPT-NIR_data}
            \end{table}
            
            \begin{table}[h]
                \centering
                \footnotesize
                \setlength{\tabcolsep}{2.4pt}
                \renewcommand{\arraystretch}{1.5}
                \caption{The power-law index $\alpha$ of afterglow of GRB230307A and temporal evolution of AT 2023vfi, which derived from the SED fitting of multi-wavelength observations at different times. $\Delta t$ is the time since GRB occurrence and errors represent the $1\sigma$ uncertainties.}
                \begin{tabular}{*{5}{c}}
                    \toprule 
                    {\bf $\Delta t$} & {\bf $T_{\rm eff}$} & {\bf $L_{\rm bol}$} & {\bf $R_{\rm ph}$} & $\alpha$\\
                    {(day)} & {(K)} & {($\rm erg/s$)} & {(cm)} & \\
                    \midrule
                    1.2 & $6853^{+1498}_{-799}$ & $7.68^{+3.13}_{-2.81} \times 10^{41}$ & $6.99^{+3.85}_{-2.44} \times 10^{14}$ & \multirow{4}{*}{$-1.77^{+0.05}_{-0.07}$}\\
                    1.8 & $5392^{+823}_{-768}$ & $4.34^{+1.30}_{-1.50} \times 10^{41}$ & $8.48^{+2.58}_{-2.49} \times 10^{14}$ &\\
                    2.4 & $3121^{+963}_{-1968}$ & $1.77^{+1.88}_{-0.84} \times 10^{41}$ & $1.62^{+0.33}_{-1.68} \times 10^{15}$ &\\
                    7.4 & $1712^{+780}_{-273}$ & $8.37^{+6.36}_{-5.50} \times 10^{40}$ & $3.70^{+4.58}_{-2.28} \times 10^{15}$ &\\
                    \midrule
                    28.9 & $588^{+19}_{-13}$ & $5.00^{+0.21}_{-0.19} \times 10^{39}$ & $7.65^{+0.48}_{-0.36} \times 10^{15}$ & \multirow{2}{*}{$-1.63^{+0.06}_{-0.05}$}\\
                    61.4 & $541^{+90}_{-107}$ & $6.51^{+4.34}_{-1.39} \times 10^{38}$ & $3.26^{+2.14}_{-1.63} \times 10^{15}$ & \\
                    \bottomrule
                \end{tabular}
                \label{tab:time_evolution}
            \end{table}
            To reduce the Swift X-Ray Telescope (XRT) data, we utilize the tools integrated within HEASOFT v6.33.2 to extract spectra from photon event files \citep{HEAsoft}. For the reduction and analysis of XMM-Newton data, we employ the Science Analysis System (SAS) v21.0.0 to obtain the spectra \citep{SAS}. A total of five X-ray spectra are obtained at different times, spanning from 1 day to 37 days post-burst. The ultraviolet, optical and near-infrared data close to each X-ray observation are collected from \cite{Levan_2024} and \cite{Yang_2024}, as shown in Table \ref{tab:UV-OPT-NIR_data}. To use the Xspec tool within HEASOFT to fit the SED from optical to X-ray, the photometric data are converted to a \texttt{.pha} file by ftflx2xsp tool within HEASOFT to input into Xspec. The SED models for the afterglow and KN are adopted as a power-law function and the black-body emission, respectively. To account for the Galactic foreground extinction and absorption, the fitting model in Xspec is set as \texttt{redden*phbabs*zphbabs*(powerlaw+bbody)}, with the parameters redshift, Galactic extinction factor and Galactic hydrogen column density fixed at: $z=0.065$, $E(B-V)=0.076$, $n_{\rm H}=1.26\times 10^{21} {\rm cm}^{-2}$ \citep{2011ApJ...737..103S,2013MNRAS.431..394W}, based on the occurrence location of GRB 230307A \citep{Levan_2024}. \par
            The fitting results are shown in Table \ref{tab:time_evolution}, where the parameter $\alpha$ is the power-law index of afterglow, and the photosphere radius $R_{\rm ph}$ at each time is derived from the fitting parameters of effective temperature $T_{\rm eff}$ and the bolometric luminosity $L_{\rm bol}$. The temporal evolution of the black-body component is consistent with the features of KNe including rapid decline and fast color evolution from blue to red. In addition, the luminosity of this component is similar to that of KN AT 2017gfo \citep{Villar_2017,Waxman_2018,metzger_kilonovae_2020}, supporting the identification of this component as originating from a KN. Our results are also consistent with the results presented in Extended Data Table 1 in \cite{Yang_2024}. With the broad near-infrared coverage and high sensitivity of JWST, the bolometric luminosity of KNe at late stages ($\sim60$ days post-burst) could be measured for the first time. \par
        \subsection{\textit{r}-process nucleosynthesis} \label{method.2} 
            As an important energy source of KN, the abundances of \textit{r}-process elements are the basic factor for the luminosity and evolution of KNe. In this section, we introduce the method to simulate the abundances of ejecta during BNS mergers. \par
\begin{figure}[ht]
                \centering
                {\includegraphics[scale=0.55]{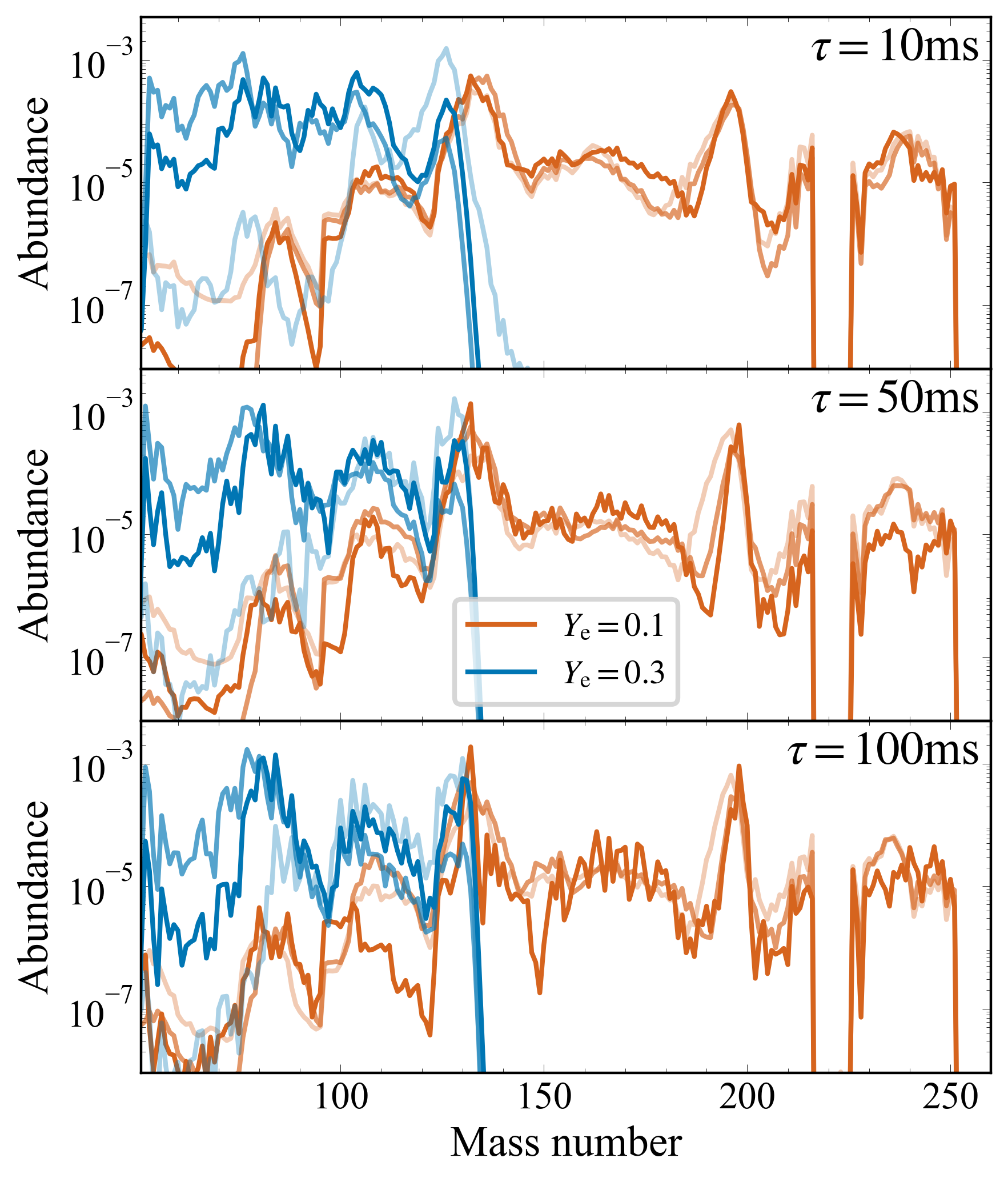}}
                \caption{The abundance distributions at 1e+9 s simulated by SkyNet nucleosynthesis code with various initial conditions. The initial temperature $T_{\rm ini}$ is fixed to 6 GK. The entropy is selected as (10, 50, 100, 200) $k_{\rm B}\,{\rm baryon}^{-1} $. For the lines with same color, the higher transparency represents the larger entropy set.}
                \label{fig:Abundances}
            \end{figure}

            In this work, we adopt a two-component KN model, including lanthanide-rich and lanthanide-poor ejecta, also named as red and blue components \citep{Villar_2017,metzger_kilonovae_2020}, respectively. For lanthanide-rich ejecta, following the method in \cite{Chen_2023a} and \cite{Chen_2024}, we derive the abundances of lanthanide-rich ejecta during BNS mergers by \textit{r}-process nucleosynthesis simulation. Specifically, we employ the nuclear reaction network code SkyNet\footnote{\href{https://jonaslippuner.com/research/skynet/}{https://jonaslippuner.com/research/skynet/}} for simulation \citep{Lippuner_2015,Lippuner_2017}. SkyNet is a general-purpose nuclear reaction network specifically designed for \textit{r}-process nucleosynthesis calculations, incorporating over 140,000 nuclear reactions and evolving the abundances of 7,843 nuclides, from free neutrons and protons up to $^{377}{\rm Cn(Z=112)}$. In our simulation, we update the nuclear mass data using the latest nuclear database from AME2020 \citep{Wang_2021}. In addition, we fit and calculate the neutron capture rates at various temperatures using the results from the nuclear reaction code TALYS \citep{2008A&A...487..767G}. The system is initially assumed to be in nuclear statistical equilibrium. It evolves according to specified initial conditions, including temperature $T_{\rm ini}$, entropy $s$, expansion timescale $\tau$, and electron fraction $Y_{\rm e}$, which serve as input parameters for SkyNet. Figure \ref{fig:Abundances} shows the final abundance distributions at $10^9$ s for various initial conditions. The electron fraction $Y_{\rm e}$, which reflects the neutron richness of the environment, primarily dominates the synthesis of \textit{r}-process elements. Higher entropy leads to lower production of lanthanides and actinides due to the inverse relationship between entropy and initial density, resulting in reduced neutron flux and density at higher initial entropy. The expansion timescale, which determines how fast the density decreases during nuclear burning, has a relatively minor impact on \textit{r}-process element synthesis compared with electron fraction and entropy. A slower expansion timescale allows the ejecta to return to nuclear statistical equilibrium more rapidly, resulting in an increased electron fraction and a reduced production of lanthanides and actinides \citep{Lippuner_2015}. \par
            To simulate abundances of lanthanide-rich ejecta produced during BNS mergers, the initial conditions are set based on the numerical relativity simulation presented in \cite{Radice_2018}. For GRB 230307A, the information of masses of the BNS system is extremely limited due to the lack of GW observation. For simplicity, we use a BNS merger system with a mass of $(1.4+1.4)M_{\odot}$ and adopt four distinct equations of state (EoSs): BHBlp\_M140140\_LK, DD2\_M140140\_LK, LS220\_M140140\_LK, and SFHo\_M140140\_LK. The EoSs of BHB$\Lambda\phi$, DD2, LS220 and SFHo support cold, nonrotating maximum NS mass and $R_{1.4}$ combinations of (2.11 $M_\odot$, 13.2 km), (2.42 $M_\odot$, 13.2 km), (2.06 $M_\odot$, 12.7 km) and (2.06 $M_\odot$, 11.9 km), respectively \citep{Typel_2010,Hempel_2010,Steiner_2013,Banik_2014}. These predicted maximum NS masses and radii are consistent with current astrophysical constraints \citep{Radice_2018}. EoSs with smaller $R_{1.4}$ are referred as ``softer", while those with larger $R_{1.4}$ are referred as ``stiffer". Additionally, we include an asymmetric system with a mass of $(1.2 + 1.4) M_{\odot}$ in our calculations, labeled as BHBlp\_M140120\_LK. The initial conditions for these binary systems, including $Y_{\rm e}$, $s$, $v_R$, are listed in Table \ref{tab:BNS_model}. Following the method in \cite{Chen_2023a}, the expansion timescale is $\tau \approx \frac{cRe}{3v_R} $, where $e \simeq 2.718$ is Euler’s number and $v_R$ is the velocity measured on a sphere with a coordinate radius $R = (300G/c^2)M_\odot \approx 443\,{\rm km}$. The derived expansion timescales for these systems are listed in Table \ref{tab:BNS_model}. \par
\begin{figure}[ht]
                \centering
                {\includegraphics[scale=0.48]{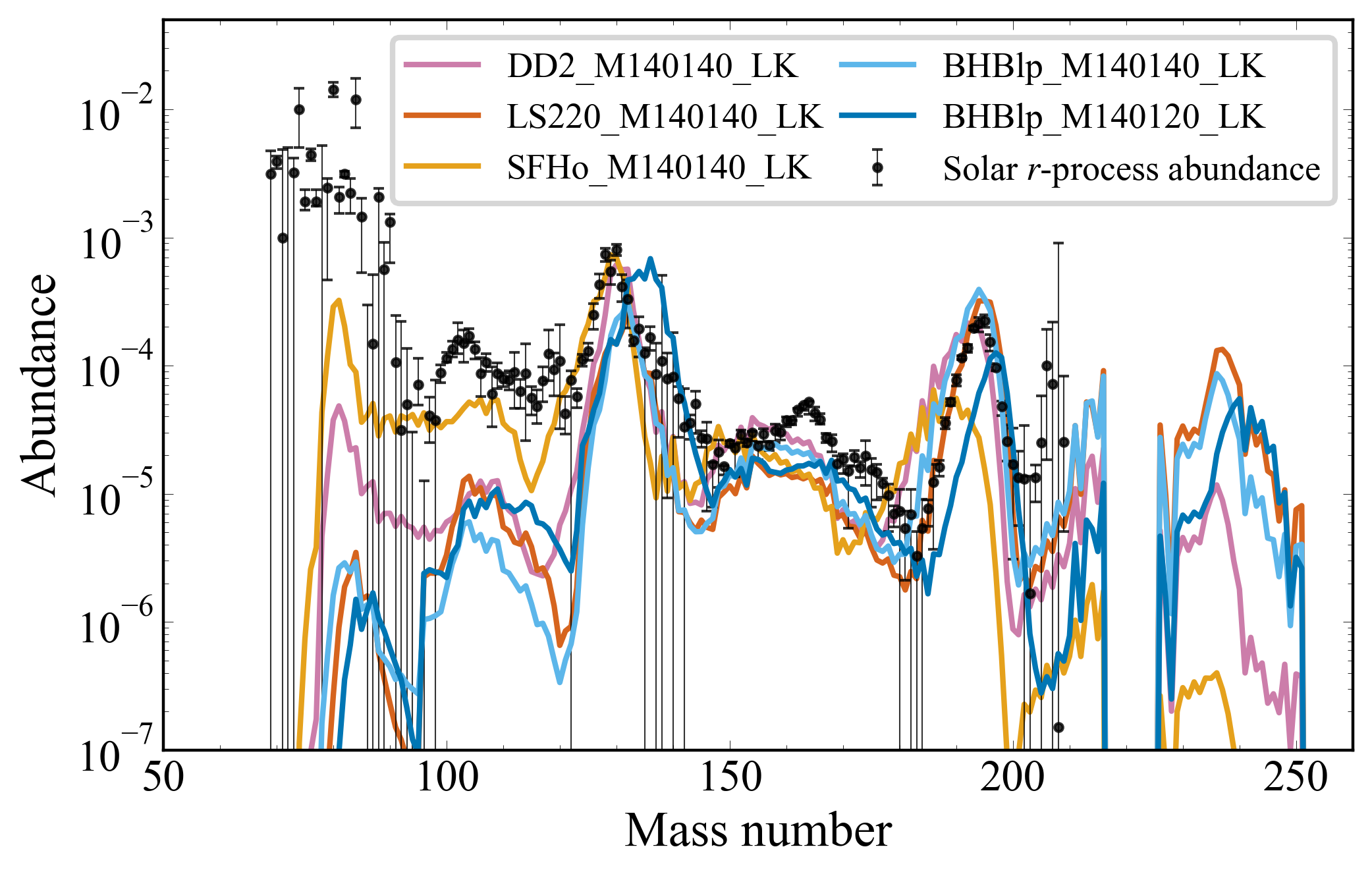}}
                \caption{The abundance distributions at 1e+9 s simulated by SkyNet nucleosynthesis code for the models with different EoSs. The black dots represent solar \textit{r}-process element abundances from \cite{Arnould_2007}.}
                \label{fig:EoSs_abundances}
            \end{figure}

            The abundance distributions simulated by SkyNet for these systems are shown in Figure \ref{fig:EoSs_abundances}, with black dots representing the solar \textit{r}-process abundance measurements from \cite{Arnould_2007}. Regardless of the EoSs, these BNS systems robustly produce second- and third-peak elements resembling the solar \textit{r}-process abundance, e.g., with $125\lesssim A \lesssim 145$ and $185\lesssim A \lesssim 210$, respectively. For the model BHBlp\_M140120\_LK with unequal mass, the abundances of actinide elements produced are relatively lower than the symmetrical system BHBlp\_M140140\_LK. For the $(1.4+1.4)M_{\odot}$ BNS system with different EoSs, the produced abundance of heavy elements is sensitive to the EoSs, particularly for light and superheavy \textit{r}-process elements, but with a complicated relationship: e.g., the more lanthanide and actinide elements are synthesized in the softer model LS220\_M140140\_LK and the stiffer model BHBlp\_M140140\_LK. \par
            In addition to the lanthanide-rich ejecta, the lanthanide-poor ejecta is a portion of ejecta matter unbound from a BNS merger, which contains a lower neutron abundance ($Y_{\rm e} \gtrsim 0.3$). Due to the environment’s low neutron density, the lanthanide-poor ejecta lacks lanthanide group elements and exhibits a lower opacity, corresponding to the blue color and early lightcurve of KNe \citep{metzger_kilonovae_2020}. Based on nucleosynthesis calculations for high $Y_{\rm e}$, as shown in Figure \ref{fig:Abundances}, we adopt a solar-like \textit{r}-process abundance for the lanthanide-poor ejecta in the mass number range of $72 \leq A \leq 140$.
            
            \begin{table}[t]
                \centering
                \small
                \renewcommand{\arraystretch}{1.5}
                \caption{The BNS simulation models and corresponding parameters used in this work. The expansion timescales $\tau$ are derived from ejecta velocity $v_R$ measured on a sphere with a coordinate radius $R = (300G/c^2)M_\odot \approx 443\,{\rm km}$.}
                \begin{tabular}{*{5}{c}}
                \toprule 
                Model & ${Y_{\rm e}}$ & $s$ & $v_R$ & $\tau$\\
                & & $(k_{\rm B}/{\rm baryon})$ & $(c)$ & $({\rm ms})$\\
                \midrule
                DD2\_M140140\_LK & 0.17 & 22 & 0.22 & 6.08\\
                LS220\_M140140\_LK & 0.14 & 16 & 0.17 & 7.87\\
                SFHo\_M140140\_LK & 0.19 & 37 & 0.35 & 3.82\\
                BHBlp\_M140140\_LK & 0.15 & 18 & 0.17 & 7.87\\
                BHBlp\_M140120\_LK & 0.11 & 13 & 0.16 & 8.36\\
                \bottomrule
                \end{tabular}
                \label{tab:BNS_model}
            \end{table}
        \subsection{Heating rate} \label{method.3}
            In this section, we present the calculation of energy release and heating rates of nuclear reactions with the input of \textit{r}-process element abundances. \par
            Following the nucleosynthesis of the \textit{r}-process elements, a substantial amount of energy is released fueling the KN by radioactive decay. The decay products carry the radioactive energy and heat the ejecta material via interactions. To calculate the energy release and heating rate, we adopt the model presented in \cite{hotokezaka_radioactive_2020} and modify the code\footnote{\href{https://github.com/hotokezaka/HeatingRate}{https://github.com/hotokezaka/HeatingRate}} to also accommodate non-solar abundance distributions. This model accounts for radioactive power from three decay channels of $\beta$-decay, $\alpha$-decay and spontaneous fission. It also incorporates the thermalization process of each decay product to derive the heating rates. The total heating rates for the three decay channels are as follows:
            \begin{equation} \label{eq:Qth_total}
                \begin{aligned}
                    \dot{Q}(t) = \sum_{\rm p}\dot{Q}_{\rm p,th}(t) = \sum_{\rm p}f_{\rm p}(t)\dot{q}_{\rm p}(t), 
                \end{aligned}
            \end{equation}
            where ${\rm p} \in [\alpha,\beta,\gamma,{\rm fission\,fragments}]$, $f_{\rm p}(t)$ is the heat deposition fraction, $\dot{q}_{\rm p}(t)$ is the total energy release carried by each kind of decay products:
            \begin{equation} \label{eq:q}
                \begin{aligned}
                    \dot{q}_{\rm p}(t) = N_{\rm A} \sum_{i} \tau_{i}^{-1} e^{t/\tau_{i}} Y_{i} E_{{\rm p},i}, 
                \end{aligned}
            \end{equation}
            where $N_{\rm A}$ is Avogadro’s number, $\tau_i$ is the mean lifetime of the nuclide, $E_{{\rm p},i}$ is the energy released per decay chain, and $Y_i$ is the number fraction of a parent nuclide per mole. The energy release and mean lifetime data are adopted from the Evaluated Nuclear Data File library \citep[ENDF/B-VII.1,][]{CHADWICK20112887}. \par
            The thermalization process is calculated separately for charged particles and $\gamma$-ray photons. For charged particles, including $\alpha$ particles, electrons, and fission fragments, the heating rate is expressed as:
            \begin{equation} \label{eq:heating_charged}
                \begin{aligned}
                    \dot{Q}_{\rm charged,th}(t) = \sum_{i} \int_{t_{0,i}}^{t} dt' \beta K_{\rm st}(E_{i,0};t',t) \rho(t) \frac{N_i(t')}{\tau_i},
                \end{aligned}
            \end{equation}
            where $N_i(t)$ is the number density of a radioactive element $i$, $\beta$ is the velocity of charged particles, $K_{\rm st}$ is the stopping cross section per unit mass, and $\rho(t)$ is the density. The term of $\beta K_{\rm st}(E_{i,0};t',t) \rho(t)$ describes the collision energy loss per unit time, where $\beta K_{\rm st}(E_{i,0};t',t)$ is obtained by solving the time evolution of the kinetic energy of monoenergetic charged particles for a given initial energy, $E_{i,0}$, and injected time $t'$. For $\gamma$-ray photons, the heat deposition fraction is:
            \begin{equation} \label{eq:heating_charged}
                \begin{aligned}
                    {f}_{\rm \gamma}(t) \approx 1-\exp(-\tau_{\gamma,_{\rm eff}}) = 1-\exp({-(t/t_0)}^2),
                \end{aligned}
            \end{equation}
            where $\tau_{\gamma,{\rm eff}}$ is effective optical depth of $\gamma$-ray photons. The time scale $t_0$ is defined by the time $\tau_{\gamma,{\rm eff}}=1$, and is estimated by:
            \begin{equation} \label{eq:time_scale}
                \begin{aligned}
                     t_0 \approx &2.3\,{\rm day} \left(\frac{C_\Sigma}{0.05} \right)^{0.5} \left(\frac{M_{\rm ej}}{0.05 M_\odot} \right)^{0.5}\\
                     & \left(\frac{v_{\rm min}}{0.1c} \right)^{-1} \left(\frac{\kappa_{\gamma,{\rm eff}}}{0.07 {\rm cm^2g^{-1}}} \right)^{-0.5},
                \end{aligned}
            \end{equation}
            where $M_{\rm ej}$ represents the ejecta mass, $v_{\rm min}$ is the minimum velocity of ejecta, $\kappa_\gamma,{\rm eff}$ is the effective opacity for $\gamma$-rays, the parameter ${C_\Sigma}$ depends on the structure of the ejecta. \par
        \subsection{KN model} \label{method.4}
            The KN lightcurve is derived using a classical semi-analytic kilonova model with two components and spherically symmetric geometry, developed in the literature \citep{Li_1998,Metzger_2010,2010ApJ...717..245K,Piran_2013,Kasen_2015}. The model is described as follows: the merger ejecta is divided into $N$ spherical shells, with each shell characterized by its mass $m_i$, a grey opacity $\kappa_{i}$, and an expansion velocities $v_i$, where $v_1 = v_{\rm min}$ and $v_N = v_{\rm max}$. The radial density of the merger ejecta is assumed as a power-law density profile: 
            \begin{equation}
                \begin{aligned}
                    \rho(v_i, t) = \rho_0(t) \left(\frac{v_i}{v_{\rm min}}\right)^{-n},
                \end{aligned}
                \label{density}
            \end{equation}
            where $n$ is set as a fiducial value of 3 according to \cite{metzger_kilonovae_2020}, $\rho_0(t)$ is the density of the innermost layer, and $\rho_0(t)$ is obtained given a total ejecta mass $M_{\rm ej}$:
            \begin{equation}
                \begin{aligned}
                    M_{\rm ej} = \int_{R_1}^{R_N}4\pi\rho(v_i,t)r^2dr,
                \end{aligned}
            \end{equation}
            where $R_i = v_i t$ is the radius of each layer. The radiation is related to the internal energy $E_i$ of each layer, of which the evolution is described by the first law of thermodynamics of radiation dominated gas:
            \begin{equation} \label{eq:energy_evolution}
                \begin{aligned}
                    \frac{dE_i}{dt} = -\frac{E_i}{t} + \dot{Q_i}(t)-L_{i}(t)\; {\rm for}\;i=1,...,N,
                \end{aligned}
            \end{equation}
            where $\dot{Q}_i(t) = m_i\dot{Q}(t)$ is the heating rate of the each mass layer. The radiation luminosity $L_{i}$ of each shell can be estimated by:
            \begin{equation}
                \begin{aligned}
                    L_i = \frac{f_{{\rm esc},i}E_i}{t_{{\rm lc},i} + \min(t_{{\rm diff},i},t)},
                \end{aligned}
            \end{equation}
            where $t_{{\rm lc},i}=v_i t/c$  is the light-crossing time and $t_{{\rm diff},i}=\tau_i v_i t/c$ is the photon diffusion time scale. The optical depth $\tau_i$ of the ith shell is:
            \begin{equation}
                \begin{aligned}
                    \tau_i(t) = \int_{R_i}^{\infty} \kappa(r) \rho(r) dr.
                \end{aligned}
            \end{equation}
            The parameter $f_{{\rm esc},i}$ is the energy escape fraction in each shell, which is estimated by:
             \begin{equation}
                \begin{aligned}
                    f_{{\rm esc},i} \approx {\rm erfc} \left(\sqrt{\frac{t_{{\rm diff},i}}{2t}}\right),
                \end{aligned}
            \end{equation}
            where erfc is the complementary error function. Finally, the total luminosity of the KN is $L_{\rm bol}=\sum_{i}L_i$.
            The two components of ejecta are defined by radially varying opacity: 
            \begin{equation}
                \begin{aligned}
                    \kappa(v_{i},t) = 
                        \begin{cases} 
                        \kappa_{\rm low} & \text{if } v_{i} \geq v_{\kappa} \\
                        \kappa_{\rm high} & \text{if } v_{i} < v_{\kappa}, 
                        \end{cases}
                \end{aligned}
            \end{equation}
            where $\kappa_{\rm low}$ and $\kappa_{\rm high}$ represent the opacity of lanthanide-poor and lanthanide-rich ejecta, respectively. In geometry, the lanthanide-poor ejecta is located in the outer layers of the ejected material with higher velocity, while the lanthanide-rich ejecta is in the inner layers. \par
            In our calculations, we set $\kappa_{\rm low}=0.5\,{\rm cm^{2} g^{-1}}$ according to the fitting results in \cite{Yang_2024}. The opacity of lanthanide-rich ejecta is set as $\kappa_{\rm high}=10\,{\rm cm^{2} g^{-1}}$ based on the best-fit model for $\sim29$ days JWST spectrum in \cite{Gillanders_2023}. The two values of opacity are also consistent with opacity ranges for lanthanide-poor and lanthanide-rich ejecta, respectively \citep{10.1093/mnras/staa1576}. For ejecta velocity, according to numerical simulations for dynamical ejecta \cite{2013ApJ...773...78B,2016PhRvD..93l4046S,dietrich_modeling_2017,Radice_2018}, the velocity range is $\sim0.1-0.4\,c$. Additionally, based on the fitting results for the observations AT 2017gfo and AT 2023vfi \citep{Villar_2017, Gillanders_2023, Yang_2024}, the velocities of lanthanide-rich and lanthanide-poor ejecta are in range of $\sim0.1-0.2c$ and $\sim0.2-0.3c$, respectively. Therefore, we set the velocity range and velocity threshold as: $v_{\rm min} = 0.1c, v_{\rm max}=0.4c, v_{\kappa}=0.18c$. Based on the density profile in Equation \ref{density}, the mass fraction of lanthanide-rich ejecta is $\sim0.4$.
    \section{Results and Discussions} \label{results}
        We present the calculated heating rates and lightcurves for different models in Section \ref{results.1}. In Section \ref{results.2}, we apply the results to AT 2023vfi to constrain the abundances of the superheavy elements.
        \subsection{Heating rate} \label{results.1}
\begin{figure}[ht]
                \centering
                {\includegraphics[scale=0.48]{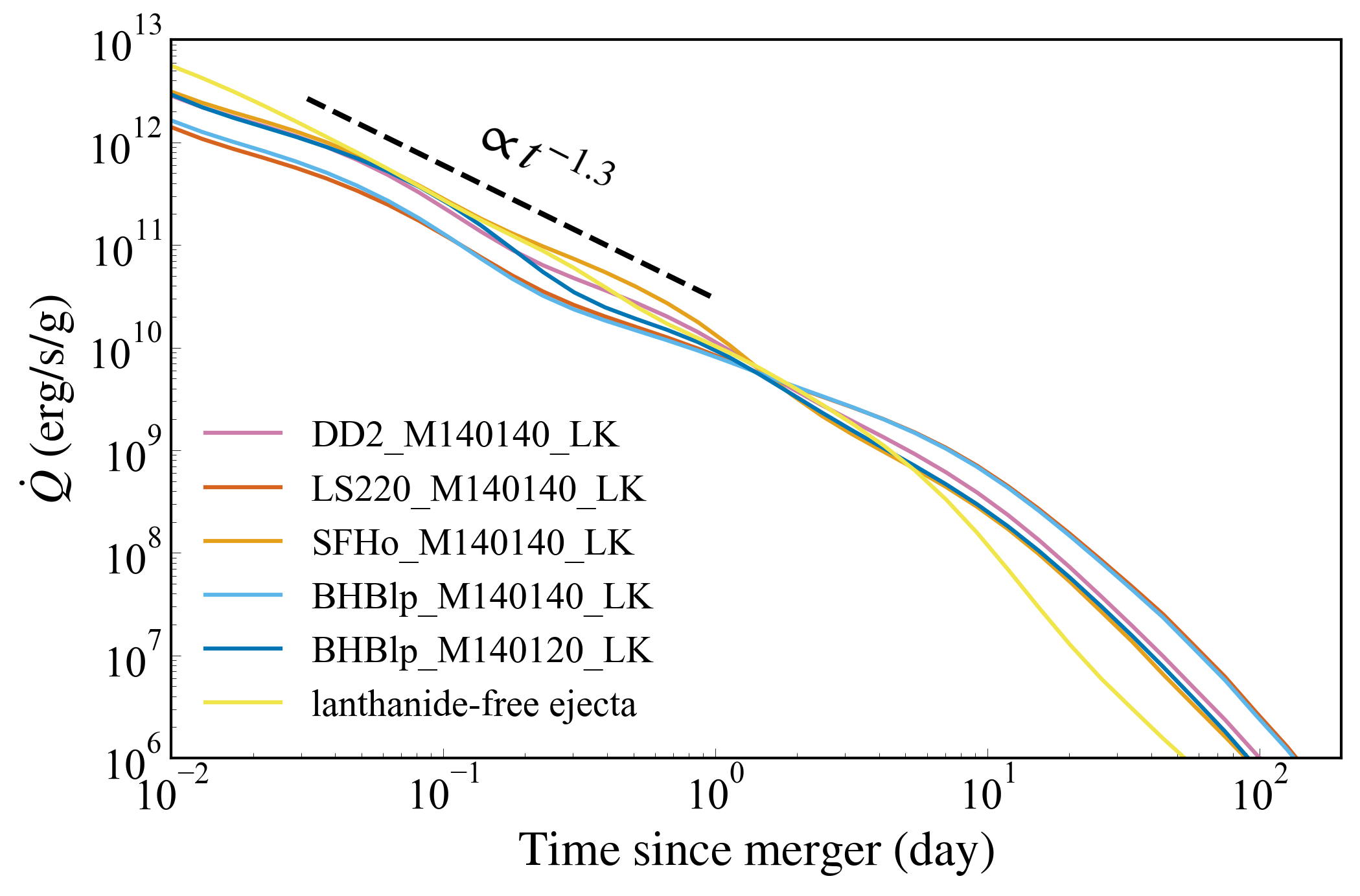}}
                \caption{The heating rates of lanthanide-poor and lanthanide-rich ejecta of different models. The dashed black line represents a power-law function with index of -1.3.}
                \label{fig:heating_rate}
            \end{figure}

\begin{figure}[ht]
                \centering
                {\includegraphics[scale=0.48]{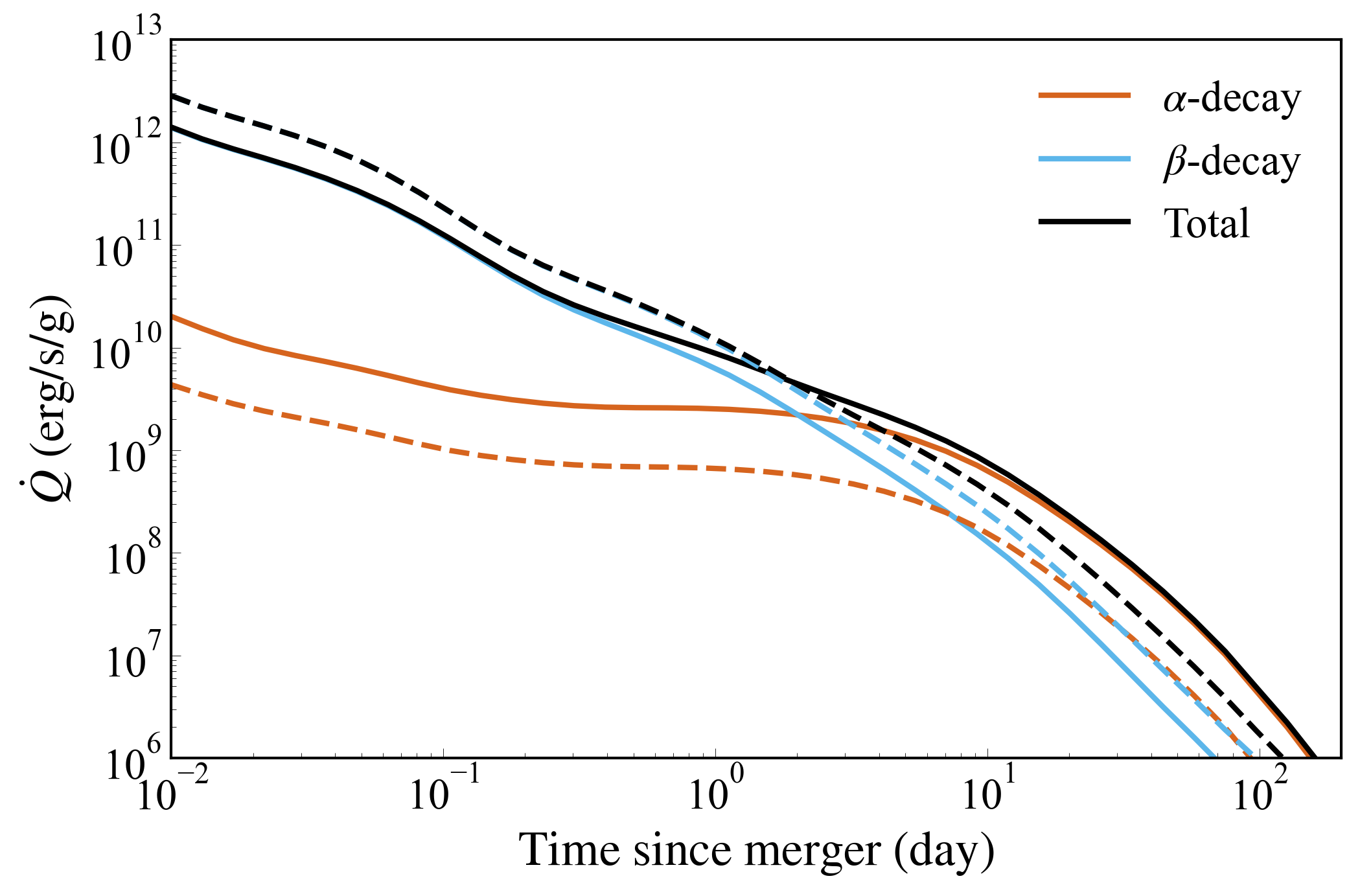}}
                \caption{The heating rates produced by the $\beta$-decay and $\alpha$-decay of \textit{r}-process elements for lanthanide-rich ejecta. The solid and dashed lines represent LS220\_M140140\_LK and DD2\_M140140\_LK, respectively.}
                \label{fig:heating_rate_component}
            \end{figure}

             By the methods introduced in Section \ref{method.2}, the heating rates are calculated for lanthanide-poor ejecta and lanthanide-rich ejecta of different models listed in Table \ref{tab:BNS_model}. For these models, the masses of ejecta are all set as $M_{\rm ej}=0.05M_\odot$ in our calculation. The results of heating rates are shown in Figure \ref{fig:heating_rate}. For all models, the early evolution of the heating rate can be well described by a power-law function with an index of $\sim$-1.3, consistent with the results in the literature \citep{Korobkin_2012,hotokezaka_2017}. However, at later epochs, heating rates of BHBlp\_M140140\_LK and LS220\_M140140\_LK decline more slowly compared to others. Such a difference is attributed to the change in the energy source from different radioactive decay channels. Figure \ref{fig:heating_rate_component} shows the heating rates from $\beta$-decay and $\alpha$-decay for LS220\_M140140\_LK and DD2\_M140140\_LK. In the first few days, the heating rate is mainly contributed by the $\beta$-decay of various \textit{r}-process elements. The combined energy release from nuclides with different half-lives leads to the observed power-law behavior. At later times, as shown in Figure \ref{fig:heating_rate_component}, the contribution of $\alpha$-decay gradually increases and becomes dominant in LS220\_M140140\_LK, corresponding to the slower decline for both BHBlp\_M140140\_LK and LS220\_M140140\_LK. \par
            
\begin{figure}[ht]
                \centering
                {\includegraphics[scale=0.48]{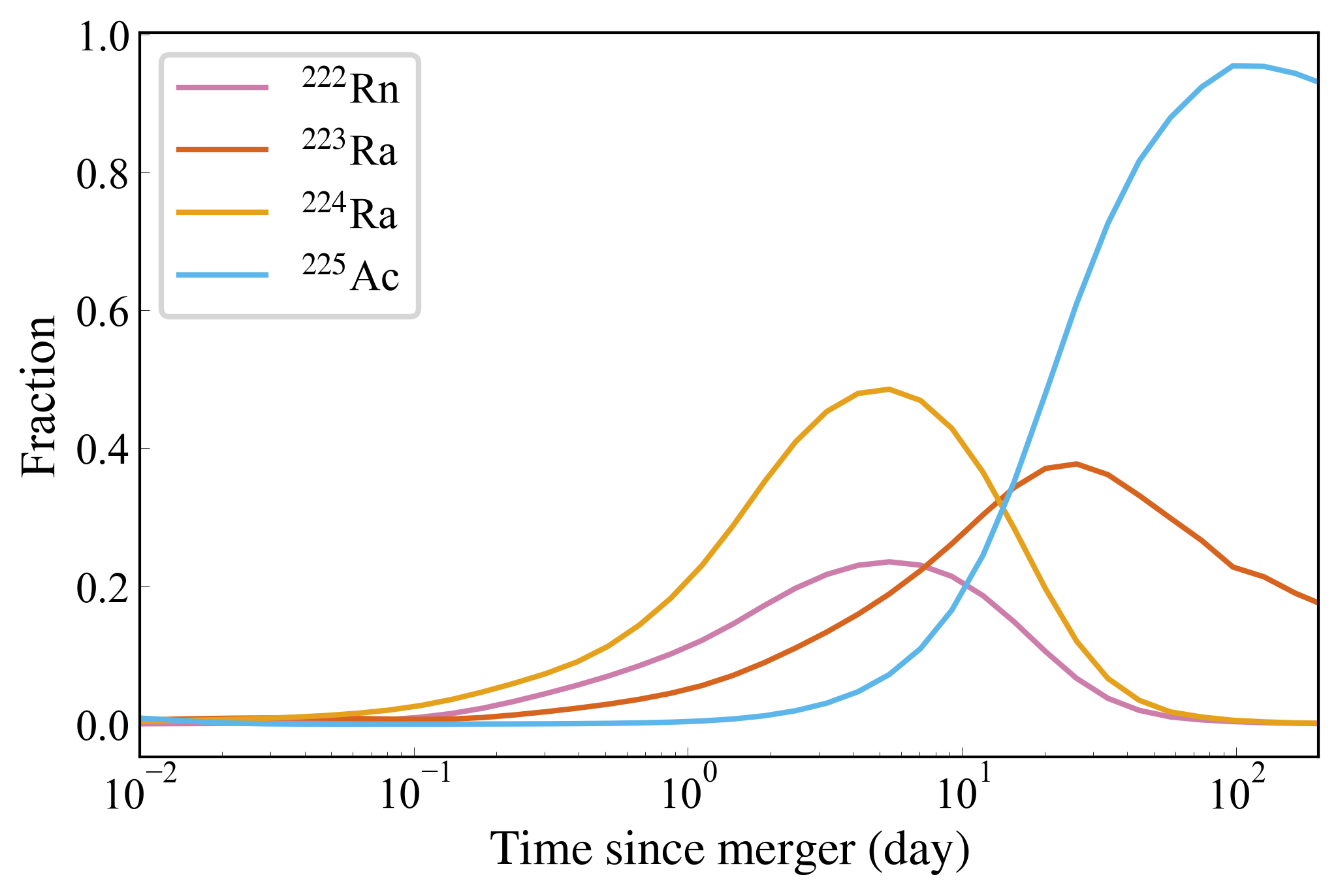}}
                \caption{The fractions of heating rate for the superheavy elements: $^{222}$Rn, $^{223}$Ra, $^{224}$Ra and $^{225}$Ac, in the case of LS220\_M140140\_LK.}
                \label{fig:alpha_fraction}
            \end{figure}

             According to \cite{Wu_2019}, at later stages, the $\alpha$-decay energy is primarily contributed by four superheavy nuclides: $^{222}$Rn, $^{223}$Ra, $^{224}$Ra and $^{225}$Ac. Heating rate fractions of the four elements for LS220\_M140140\_LK are shown in Figure \ref{fig:alpha_fraction}. The half-lives of these elements are 3.8, 11.4, 3.6, 10.0 days, and the total energies released per decay are 23.8, 30.0, 30.9, 30.2 MeV, respectively. The relatively longer half-lives and the higher energy released make these elements dominant sources of heating rate from a few days after the GRB. According to the results in Figure \ref{fig:EoSs_abundances}, the abundances of these elements produced in BHBlp\_M140140\_LK and LS220\_M140140\_LK are higher, resulting in the higher heating rates at late times compared to other models. Therefore, the late-time KN lightcurve is sensitive to these superheavy elements and can be used to estimate their abundances.\par
        \subsection{Applied to GRB 230307A} \label{results.2}
\begin{figure}[ht]
                \centering
                {\includegraphics[scale=0.48]{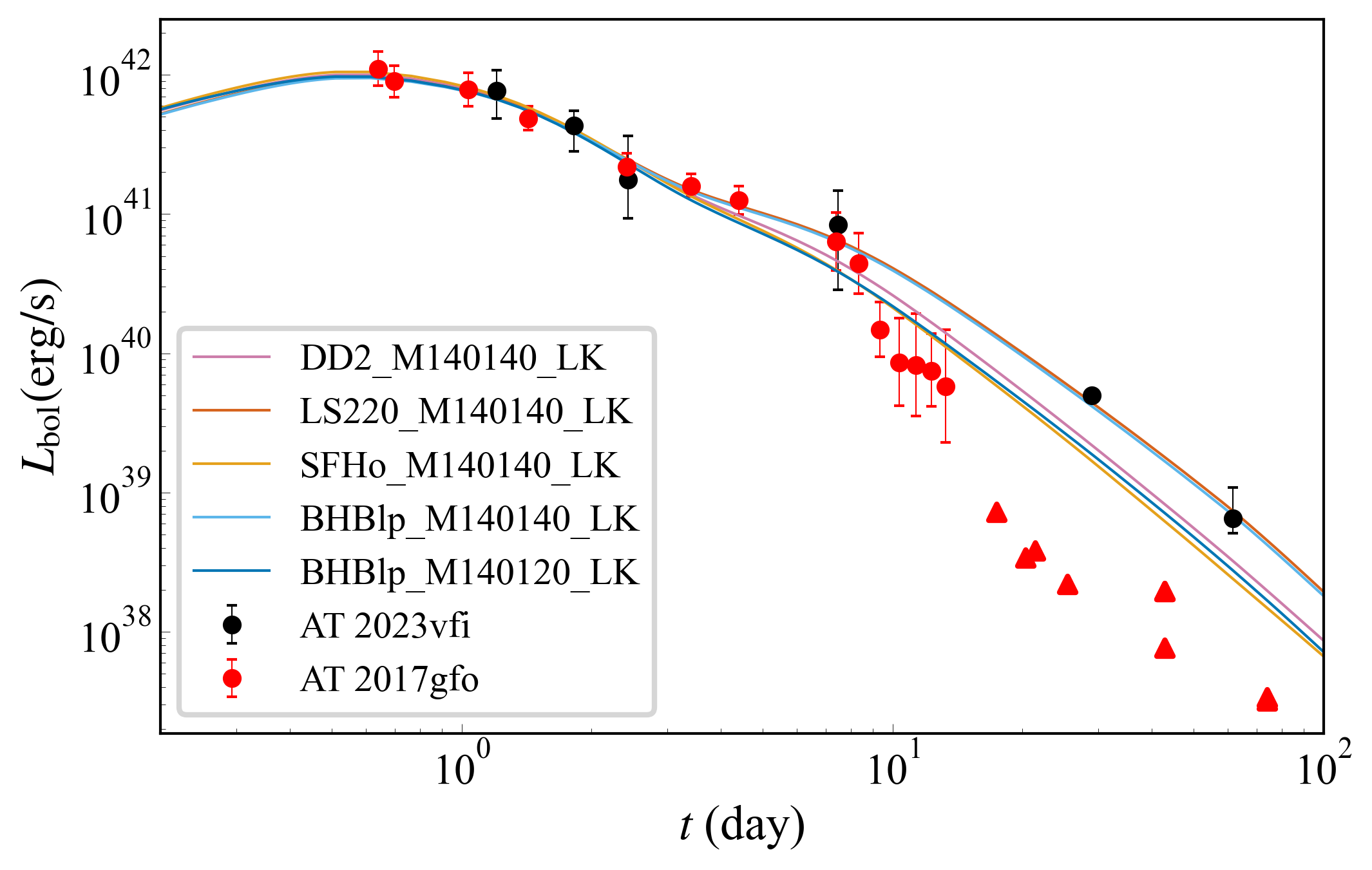}}
                \caption{The lightcurves derived from the heating rates of the BNS merger models. The dark and red dots represent the bolometric luminosity lightcurves of the KN of GRB 230307A and AT 2017gfo, respectively. The red triangles are the lower limits of bolometric luminosity for AT 2017gfo.}
                \label{fig:LC_EoSs}
            \end{figure}

            To constrain the abundances of the four superheavy elements, we calculate KN bolometric lightcurves for the five models of lanthanide-rich ejecta, using the results of heating rates from ejecta of the two components. In these calculations, the ejecta mass is set to match the early luminosity of extracted KN lightcurve presented in Table \ref{tab:time_evolution}. Specifically, the values of ejecta mass $M_{\rm ej}$ for the different models are all set as $M_{\rm ej}=0.05 M_\odot$, corresponding to $\sim0.03 M_\odot$ and $\sim0.02 M_\odot$ for lanthanide-poor and lanthanide-rich ejecta, respectively. The AT 2023vfi and model predicted bolometric lightcurves are shown in Figure \ref{fig:LC_EoSs}. For comparison, the bolometric lightcurve of AT 2017gfo in \cite{Wu_2019} is also included as represented by red dots and triangles, where triangles indicate the lower limits. We find that the bolometric lightcurves of the AT 2023vfi and AT 2017gfo are similar in the first week post-burst. This consistency suggests that AT 2017gfo and GRB 230307A are generated from a similar progenitor scenario. However, at $\sim10$ days post-burst, due to the absence of simultaneous X-ray and optical data, it remains uncertain whether AT 2023vfi exhibits a rapid decline similar to that of AT 2017gfo. About three weeks later, unlike AT 2017gfo, which only provides lower limits due to the lack of multi-band detection, two bolometric luminosity data points for the KN are measured, thanks to the high sensitivity of the JWST in the near-infrared wavelength range. \par

            The bolometric lightcurves match well with the observation of AT 2023vfi at early times for all models. However, after $\sim20$ days, only BHBlp\_M140140\_LK and LS220\_M140140\_LK are able to fit the data. For other models (e.g., DD2\_M140140\_LK and SFHo\_M140140\_LK), fitting the later data points requires an additional lanthanide-rich ejecta of $\sim0.03 M_\odot$, leading to a lanthanide-rich ejecta mass of $\sim0.05 M_\odot$ and a total ejecta mass of $\sim0.08 M_\odot$ for the two components, which is consistent with the results of the two-component KN model in \cite{Yang_2024}. \cite{Yang_2024} adopts a solar-like \textit{r}-process distribution, which lacks the superheavy elements and is similar to the models with fewer superheavy elements. For BHBlp\_M140140\_LK and LS220\_M140140\_LK, the energy from $\alpha$-decay of the superheavy elements reduce the mass of lanthanide-rich ejecta required to fit the late-time data. The less mass of 0.02 $M_\odot$ of lanthanide-rich ejecta is consistent with the dynamical ejecta mass results derived from various numerical simulations of BNS merger in literature (see Figure 5 in \cite{metzger_kilonovae_2020} and Figure 6 in \cite{dietrich_modeling_2017}). Based on this, these findings suggest that, after $\sim20$ days, the energy released by $\alpha$-decay of superheavy elements plays a significant role in powering AT 2023vfi. \par
            \begin{figure}[ht]
                \centering
                {\includegraphics[scale=0.48]{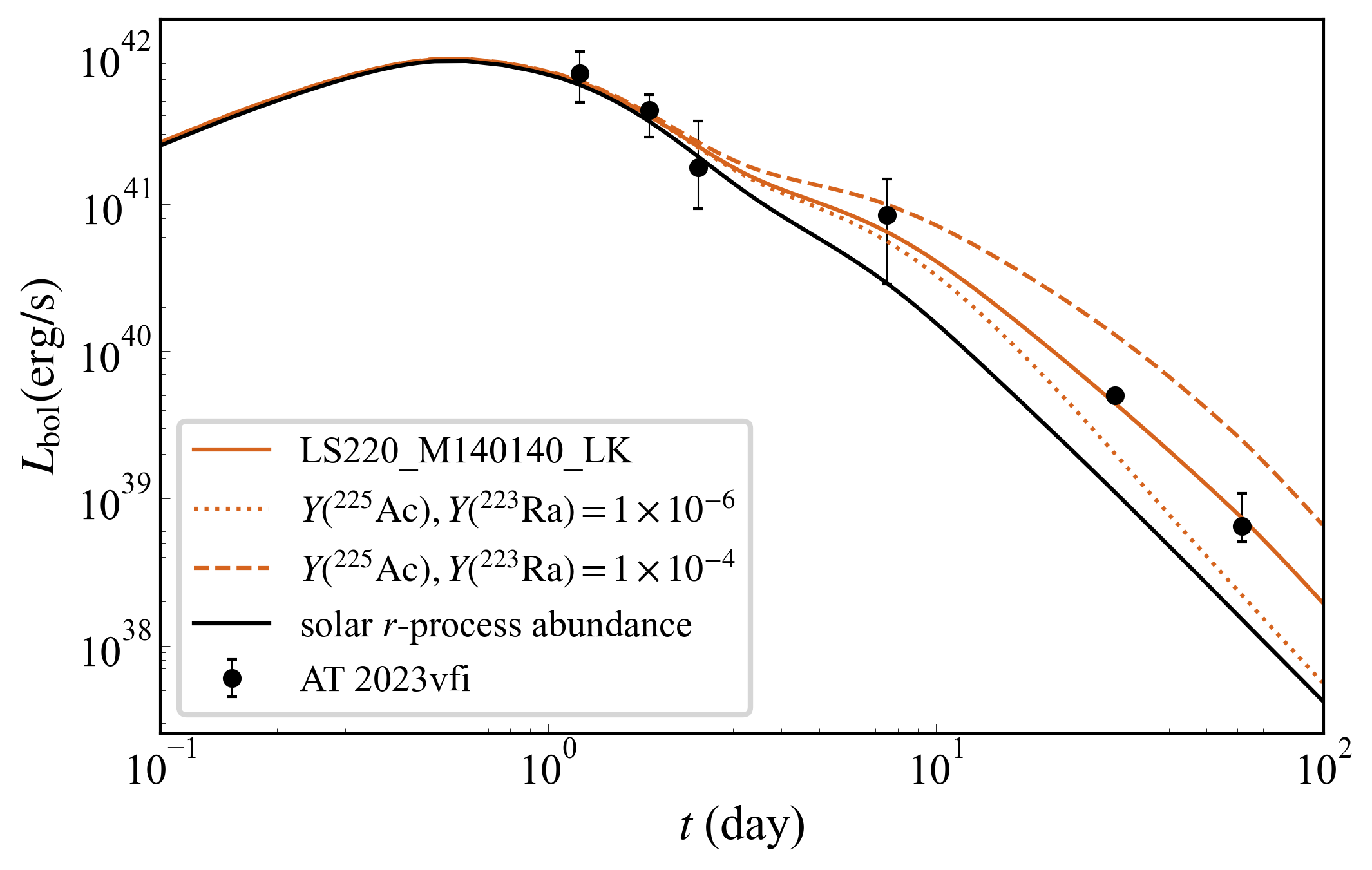}}
                \caption{Lightcurves derived from heating rates with different abundances of $^{223}$Ra and $^{225}$Ac. The black solid line is the lightcurve calculated based on the solar \textit{r}-process abundance.}
                \label{fig:LC_alpha}
            \end{figure}
            For LS220\_M140140\_LK (BHBlp\_M140140\_LK), the simulated pre-radioactive-decay abundances of $^{222}$Rn, $^{223}$Ra, $^{224}$Ra and $^{225}$Ac are: 2.1(2.2), 2.0(1.9), 3.1(2.8) and 2.6(2.2)$\times10^{-5}$, respectively, which are also consistent with the simulated abundances of $\sim2\times10^{-5}$ reported in \cite{Chen_2023a}. To investigate how the KN lightcurve changes with varying abundances of these superheavy elements, we calculate the other two lightcurves for LS220\_M140140\_LK, with abundances of both $^{223}$Ra and $^{225}$Ac set to $10^{-6}$ and $10^{-4}$, respectively. The late-time lightcurve is more sensitive to these two elements because the half-lives of both $^{223}$Ra and $^{225}$Ac are longer than 10 days. The results are shown in Figure \ref{fig:LC_alpha}, with the lightcurve corresponding to solar-like \textit{r}-process abundance for lanthanide-rich ejecta also shown as the black solid line. In our calculation, the mass number range of nuclei for solar \textit{r}-process abundance spans from 70 to 209. Due to the absence of energy contributions from the decay of the superheavy nuclei, the luminosity of solar \textit{r}-process abundance decreases more rapidly than observations. With the abundances of $10^{-6}$ and $10^{-4}$ and the mass of 0.02 $M_\odot$ of lanthanide-rich ejecta, the KN lightcurves deviate significantly from the observations after $\sim10$ days. To match with the observations, the mass of lanthanide-rich ejecta of $\sim0.05$ and $\sim0.01$ $M_\odot$ are needed for the cases of $10^{-6}$ and $10^{-4}$, respectively. When the abundances of the superheavy elements are $\geq 10^{-5}$, the less mass of lanthanide-rich ejecta is needed, which indicates that an abundance on the order of at least $\sim1\times10^{-5}$ for $^{222}$Rn, $^{223}$Ra, $^{224}$Ra and $^{225}$Ac is synthesized in the BNS merger associated with GRB 230307A. \par
            Using the estimated abundances of these elements, the quantity of other superheavy elements generated during the merger could be inferred, based on the simulated abundance ratio of the models \citep{Wu_2019}. Taking the element uranium as an example, the long-live isotopes of uranium include: $^{233}$U, $^{234}$U, $^{235}$U, $^{236}$U and $^{238}$U. According to the abundance distribution of LS220\_M140140\_LK, the abundance ratio between the uranium isotopes and $^{225}{\rm Ac}$ is $\sim0.13$. With the abundances of $\sim1\times10^{-5}$ for $^{225}{\rm Ac}$, the converted mass fraction of uranium is $\sim3.1\times10^{-2}$. Taking the mass of lanthanide-rich ejecta used in the lightcurve calculation as $\sim 0.02 M_\odot$, a total amount of $\sim6.2\times10^{-4}\,M_\odot$ is inferred to be produced during the merger. \par
        \subsection{The impact of $^{254}$Cf} \label{Cf}
            As suggested in the literature \citep{Zhu_2018,Wanajo_2018,Wu_2019}, the presence of the spontaneous fission nucleus $^{254}$Cf could potentially impact the late-time KN lightcurves due to its larger decay energy of $\sim185$ MeV and longer half-live of $\sim60.5$ days. In our calculation, abundances of $^{254}$Cf predicted by different models are within an order of $1\times10^{-9}$, which is consistent with the nucleosynthesis results of $1\times10^{-7}$ to $1\times10^{-9}$ in \citep{Chen_2023a}. The radioactive energy released by spontaneous fission from $^{254}$Cf is significantly lower compared to the energy released by $\beta$- and $\alpha$-decay. Therefore, within our adopted nuclear input, we do not find a notable contribution of $^{254}$Cf to the heating rate. \par
            \begin{figure}[ht]
                \centering
                {\includegraphics[scale=0.48]{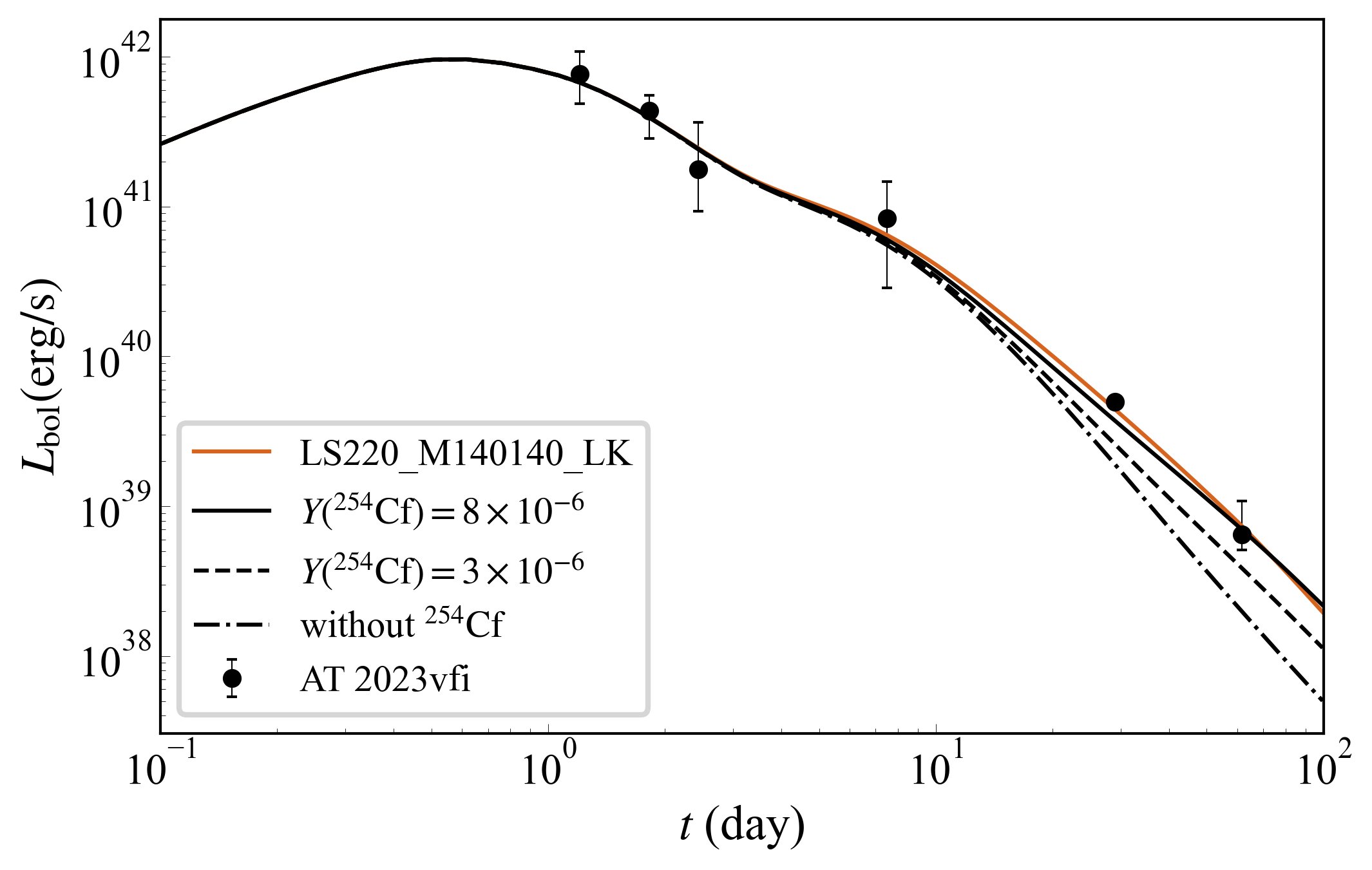}}
                \caption{The lightcurves derived from the heating rates with the varying abundances of the $^{254}$Cf.}
                \label{fig:LC_Cf}
            \end{figure}
            To explore how large the abundance of $^{254}$Cf is needed to impact late-time lightcurve significantly, we calculate the KN lightcurves with various $^{254}$Cf abundances, as shown in Figure \ref{fig:LC_Cf}. For comparison with the late-time energy released by $\alpha$-decay, for nuclide input of the lightcurves with $^{254}$Cf, we adopt the simulated abundance from LS220\_M140140\_LK for the nuclide input and set the abundances of $^{223}$Ra and $^{225}$Ac to zero. When the abundance of $^{254}$Cf increases to be $8\times10^{-6}$, the impact from the spontaneous fission is comparable to that from $^{223}$Ra and $^{225}$Ac. This finding is also consistent with the results in \cite{Zhu_2018}, where an abundance on order of $1\times10^{-6}$ for $^{254}$Cf has a notable impact to the KN lightcurve.\par
        \subsection{Other energy sources} \label{mechanisms}
            Apart from the energy released by the radioactive decay of \textit{r}-process elements, some mechanisms can also supply energy to KN and significantly affect the lightcurve, including the interaction between ejecta and CSM, shock cooling, and the energy injection from the merger remnant. Below, we discuss how these mechanisms compete with radioactive decay energy and influence our estimation of superheavy element abundances. For the ejecta–CSM interaction, this process primarily depends on the kinetic energy of the ejecta and the density of CSM, as seen in previous studies of the CSM interaction supernovae \citep{1994ApJ...420..268C,2011ApJ...729L...6C,2013ApJ...773...76C,2018ApJ...856...59L}. The ejecta–CSM interaction is applied firstly in \cite{Wang_2024a} to explain the KN emission associated with GRB 191019A. However, unlike GRB 191019A, which occurred near the nucleus of its host galaxy (within $\sim100$ parsecs in projection) in a high-density environment \citep{2023NatAs...7..976L}, the location of GRB 230307A is far from its host galaxy, with a projected distance of 38.9 kiloparsecs \citep{Levan_2024}, indicating a relatively low-density medium. Therefore, the late-time energy contribution from ejecta–CSM interaction should be negligible compared with the radioactive decay for AT 2023vfi. \par
            Shock cooling occurs when a GRB jet interacts with merger ejecta \citep{kasliwal_illuminating_2017,Piro_2018}. The main feature of shock cooling is the early-time blue light in KN lightcurve, which is similar to the thermal emission of lanthanide-poor ejecta (referred to as the blue component in the KN model). If the shock cooling mechanism significantly contributes to the energy observed in the early-time lightcurve, the ejecta mass may be overestimated, resulting a larger abundance of superheavy elements to explain the late-time lightcurve. The early detection of KN at short wavelengths within a few hours post-merger can help distinguish them \citep{arcavi_first_2018}. For AT 2023vfi, it is challenging to distinguish between shock cooling and \textit{r}-process decay based on the first measurement at $\sim 1$ day post-burst, as the afterglow dominates during this epoch. The impact of shock cooling on superheavy element constraints can be explored in future kilonovae with earlier observations. \par
            Energy injection from the merger remnant is widely discussed in literature \citep{Yu_2018,Kawaguchi_2020,Ai_2022,Wang_2023c,Ai_2024}. If the merger remnant is a long-lived magnetar, the interaction between the magnetar wind and the ejecta could enhance the KN emission. According to the latest research on engine-fed KN by \cite{Ai_2024}, in cases with relatively high spin-down luminosity of the magnetar, the KN lightcurve exhibits an early blue bump and a bluer SED. When the spin-down luminosity is relatively low, the KN lightcurve shows a substantial brightening at late epochs, and the late-time SED of KN is also bluer than the typical \textit{r}-process KN due to the continuous energy injection. For both two cases, a magnetar-enhanced KN results in a bluer SED in late stages. However, the red color of the KN revealed by JWST in the late stages indicates that the contribution from the energy injection of the merger remnant is negligible compared to \textit{r}-process radioactive decay. \par
    \section{Conclusion} \label{summary}
        This work constrains the abundances of superheavy elements of $^{222}$Rn, $^{223}$Ra, $^{224}$Ra and $^{225}$Ac, produced in the merger associated with GRB 230307A, under assumption of a BNS merger origin. To extract the the luminosity evolution of the KN, we fit the total SED contributed by the KN and afterglow, using the multi-wavelength observations in X-ray, optical and near-infrared bands. For the SED shapes, we assume a black-body emission for the KN and a power-law function for the afterglow. As a result, we obtain the evolution of the KN luminosity and temperature evolution spanning from 1 day to 61 days post-burst. \par
        To model the \textit{r}-process nucleosynthesis for lanthanide-poor and lanthanide-rich ejecta in BNS mergers, we employ the nuclear reaction network code SkyNet. Using the latest numerical relativity simulations, we simulate the \textit{r}-process nucleosynthesis for lanthanide-rich ejecta with various EoSs. To relate the energy released by radioactive decay to the KN lightcurve, the method in \cite{hotokezaka_radioactive_2020} is adopted to compute the heating rate for each nuclide. We then use a standard semi-analytic kilonova model with two components to calculate the KN lightcurve based on the heating rate. The main results of this work are summarized as follows:
        \begin{itemize}
            \item[1] The evolution of the bolometric luminosity of AT 2023vfi resembles that of AT 2017gfo in the first week post-burst, suggesting a similar progenitor scenario for both GRB 230307A and AT 2017gfo.
            \item[2] The pre-radioactive-decay abundances of $^{222}$Rn, $^{223}$Ra, $^{224}$Ra and $^{225}$Ac produced in the merger associated with GRB 230307A are constrained at least on the order of $\sim1\times10^{-5}$, which is consistent with the estimated abundances of these elements for AT 2017gfo\citep{Chen_2023a}.
        \end{itemize} \par
        The O4c observing run of LIGO/Virgo/KAGRA (LVK) gravitational wave detector network is currently underway and is planned to observe until October 2025 according to the LVK observing plans\footnote{\href{https://observing.docs.ligo.org/plan/}{https://observing.docs.ligo.org/plan/}}. The network is expected to detect BNS mergers and NS-BH mergers within detection ranges of approximately 170 and 300 Mpc, respectively \citep{abbott_prospects_2020}. Following the trigger by GW alerts, KN candidates identified by ground-based telescopes can then be monitored by JWST. In the case of serendipitous discoveries of KNe, an increasing number of wide-field telescopes are being deployed to conduct systematic surveys, including: the Vera C. Rubin observatory \citep[VRO/LSST;][]{lsst_science_collaboration_science-driven_2017,ivezic_lsst_2019}, the Wide Field Survey Telescope \citep[WFST;][]{Wang_2023,Liu_2023}, the Zwicky Transient Facility \citep[ZTF;][]{Bellm_2019,Graham_2019}. With deeper and larger surveys, more KN candidates are expected to be detected. With an expanding sample of relatively complete KN lightcurves, critical questions such as the production of heavy elements in NS-BH mergers and the relationship between heavy element nucleosynthesis and compact binary mergers are expected to be addressed. \par
    \begin{acknowledgments}
        This work is supported by the Strategic Priority Research Program of the Chinese Academy of Science (Grant No. XDB0550300), the National Natural Science Foundation of China (Grant No. 12325301, 12273035, and 12393811), the National Key R\&D Program of China (Grant No. 2023YFA1608100), and Cyrus Chun Ying Tang Foundations. J.J. acknowledges the supports from the Japan Society for the Promotion of Science (JSPS) KAKENHI grants JP22K14069.
    \end{acknowledgments}
    \vspace{5mm} 

    \software{\texttt{NumPy} \citep{harris2020array, van2011numpy}, \texttt{Astropy} \citep{robitaille2013astropy}, \texttt{matplotlib} \citep{hunter2007matplotlib}}


    
    \vspace{50mm}
    \bibliographystyle{aasjournal}
    \bibliography{draft}
    

\end{document}